\documentclass{aastex6}

\usepackage{natbib}
\usepackage{epsfig,amsmath}
\usepackage{graphicx}
\usepackage{amssymb}
\usepackage{epsfig}
\usepackage{graphics}
\usepackage{url}

\slugcomment{Received; published }
\journalinfo{Submitted to The Astrophysical Journal}

\def\arcsec{$^{\prime\prime\,}$}

\newcommand{\be}{\begin{equation}}
\newcommand{\ee}{\end{equation}}

\def\lum{erg~s$^{-1}$}

\def\smc{SMC\,X-2}

\begin{document}

\title{Propeller effect in the transient X-ray pulsar \smc}

\shorttitle{Propeller effect in the transient X-ray pulsar \smc}
\shortauthors{Lutovinov et al.}

\author{
Alexander~A.~Lutovinov\altaffilmark{1,2},
Sergey\,S.\,Tsygankov\altaffilmark{3,1},
Roman\,A.\,Krivonos\altaffilmark{1},
Sergey\,V.\,Molkov\altaffilmark{1}
and Juri\,Poutanen\altaffilmark{3,4}
}

\altaffiltext{1}{Space Research Institute, Russian Academy of Sciences, Profsoyuznaya 84/32, 117997 Moscow, Russia}
\altaffiltext{2}{Moscow Institute of Physics and Technology, Moscow region, Dolgoprudnyi, Russia}
\altaffiltext{3}{Tuorla Observatory, Department of Physics and Astronomy, University of Turku, V\"ais\"al\"antie 20, FI-21500 Piikki\"o, Finland}
\altaffiltext{4}{Nordita, KTH Royal Institute of Technology and Stockholm University, Roslagstullsbacken 23, SE-10691 Stockholm, Sweden}

\begin{abstract}
We report results of the monitoring campaign of the transient X-ray pulsar
\smc\ performed with the {\it Swift}/XRT telescope in the period of Sept 2015
-- Jan 2016 during the Type II outburst. During this event bolometric
luminosity of the source ranged from $\simeq10^{39}$ down to $several \times
10^{34}$~\lum. Moreover, we discovered its dramatic drop by a factor of more
than 100 below the limiting value of $L_{\rm lim}\simeq4\times10^{36}$~\lum,
that can be interpreted as a transition to the propeller regime. These
measurements make \smc\ the sixth pulsating X-ray source where such a
transition is observed and allow us to estimate the magnetic field of the
neutron star in the system $B\simeq3\times10^{12}$~G, that is in
agreement with independent results of the spectral analysis.
\end{abstract}

\medskip


\section{Introduction}
\label{sec:intro}

The Small Magellanic Cloud (SMC) is the satellite of the Milky Way situated
at the distance of $d\simeq62$~kpc \citep{2012AJ....144..107H}. This galaxy
is extremely rich in Be X-ray binary systems, harboring a neutron star
orbiting around an OBe companion \citep[see recent review of ][and references
therein]{2015MNRAS.452..969C}. \smc\ was discovered at an early stage of the
study of SMC \citep{1978ApJ...221L..37C} during the Type II outburst with an
X-ray luminosity in the 2--11 keV energy band of about $10^{38}$\,\lum. A
pulsating nature of the source was established during the second registered
outburst, when the pulsations with the period of $P_{\rm spin}\simeq2.37$ s
were detected with the {\it RXTE} observatory from the sky region around
\smc\ \citep{2001ApJ...548L..41C} and confirmed later using the {\it ASCA}
data \citep{2001PASJ...53..227Y}. The pulsar magnetic field was not known
until present time, but some hint at the cyclotron line detection in the
\smc\ spectrum near $\sim27$ keV was reported recently by \citet{cycline}.

The optical counterpart of \smc\ was not unambiguously identified during a
quite long time as two different stars of early spectral type are located
near the X-ray position of the source with the angular separation only
2.5\arcsec. Only recent monitoring observations with the OGLE experiment
allowed to reveal a variability of one of these stars with the period of
$P_{\rm orb}=18.62\pm0.02$ days \citep{2011MNRAS.412..391S}, that is in
agreement with periodical variations of the pulse period detected by the {\it
RXTE} and {\it Swift} observatories at $P_{\rm orb}\simeq18.4$ days
\citep{2011MNRAS.416.1556T, 2016MNRAS.458L..74L}. This periodicity was
interpreted as an orbital period in the system that, in a combination with
the pulse period of $P_{\rm spin}\simeq2.37$~s, places \smc\ in the
Be-systems region in the Corbet diagram \citep{1986MNRAS.220.1047C}.

The transient nature of a majority of Be systems is ideally suited to study
the magnetic field, to investigate luminosity-dependent accretion processes
and the geometry of the system, as well as to learn about the interaction of
the accreted matter with the neutron star magnetosphere \citep[see, e.g.,
][for reviews and current physical models]{1998A&A...338..505N,
2011Ap&SS.332....1R, 2013ApJ...777..115P,walter15}. One of the most
interesting and straightforward manifestation of such an interaction is a
transition of the accreting neutron star to the so-called propeller regime.
The physical aspects of this regime was considered by
\citet{1975A&A....39..185I}, who showed that under some conditions the
accreted matter can be stopped by the centrifugal barrier set up by the
rapidly rotating magnetosphere of the strongly magnetized  neutron star. It
should lead to the dramatic drop of the X-ray intensity of the source. The
moment of the transition from the normal accretion regime to the propeller
one depends on a combination of three physical parameters of the system --
the pulse period, magnetic moment (or magnetic field strength) of the neutron
star and the accretion rate. Because the pulse period and the accretion rate
can be derived from observations, the detection of the propeller effect
provides us with an independent estimation of the neutron star magnetic
field. This knowledge is very important as the magnetic field is one of the
fundamental parameters governing observed properties of neutron stars.

Until recently only a few cases of possible transitions into the propeller
regime in accreting millisecond and X-ray pulsars were reported in the
literature
\citep{1986ApJ...308..669S,1997ApJ...482L.163C,2001ApJ...561..924C,2008ApJ...684L..99C}.
Recently  the propeller effect was also discovered in the first pulsating
ultra-luminous X-ray source M82\,X-2 \citep{2016MNRAS.457.1101T}. This
discovery initiated a special monitoring program of transient X-ray pulsars
with the {\it Swift}/XRT telescope to search for the propeller effect in
other sources. First results of this program were published by
\cite{2016A&A...593A..16T} for two well known transient X-ray pulsars
V\,0332+53 and 4U\,0115+634, where the propeller effect was firmly
established.

In this paper we report a discovery of the propeller effect and
consequent determination of the magnetic field strength in another
transient X-ray pulsar \smc.

\section{Data analysis}
\label{sec:res}

\smc\ entered into a new outburst in the end of September 2015 and
immediately started to be monitored with the {\it Swift}/XRT telescope
\citep{2015ATel.8091....1K}. The observations (in total of around 150
individual pointings) were performed in the period from 2015 September 24  to
2016 January 25 in the Windowed Timing (WT) and Photon Counting (PC) modes
depending on the source brightness. Final products (spectrum in each
observation) were prepared using online tools provided by the UK Swift
Science Data Centre
\citep{2009MNRAS.397.1177E}\footnote{\url{http://www.swift.ac.uk/user\_objects/}}.
The spectra were grouped to have at least 1 count per bin and were fitted in
the {\sc XSPEC} package using the Cash statistic \citep{1979ApJ...228..939C}.
To avoid any problems caused by the calibration uncertainties at low
energies,\footnote{\url{http://www.swift.ac.uk/analysis/xrt/digest\_cal.php}}
we restricted our spectral analysis to the 0.7--10 keV band.

The obtained spectra in the source high state can be well fitted with a
simple powerlaw model, modified by the interstellar absorption at low
energies in the form of the {\sc phabs} model in the {\sc XSPEC} package. We
found that the hydrogen column density agrees well with the interstellar one
in this direction, therefore in the following analysis it was fixed at
$N_{\rm H}=3.4\times10^{21}$ cm$^{-2}$ \citep{LAB}. To calculate the
unabsorbed source flux the {\sc cflux} routine from the {\sc XSPEC} package
was used. In the low state the source was not detected in any single
observation. Therefore we averaged them into two groups, according to the
observational dates: the first group includes observations performed in the
second half of Dec 2015 (3 observations), just after the expected transition
to the propeller regime, the second group includes observations performed in
Jan 2016 (11 observations). Again, the source was not detected in both groups
and only upper limits to its flux were obtained based on the XRT sensitivity
curve \citep{2005SSRv..120..165B}. The log of {\it Swift}/XRT observations,
including time of observations, exposure, mode and unabsorbed flux (or upper
limits), is presented in Table\,\ref{xrt_all}.

\smc\ has been observed several times with the {\it NuSTAR} observatory that
allowed us to reconstruct its broadband spectrum in the 3--79 keV energy band
for different luminosity levels and to obtain the bolometric correction of
the flux observed by XRT in the 0.5--10 keV energy band to the flux in the
0.5--100 keV band, that can be considered as a bolometric one. The
analysis of the broad-band spectrum of SMC\,X-2 was presented by
\citet{cycline}, therefore we will not discuss it in details. Here we only
would like to mention that it can be well
described by the {\sc cutoff} model with the inclusion of the thermal
black-body component with the temperature of $\simeq1$ keV and the cyclotron
absorption line at the energies of 28--30 keV. Note, that the latter one
demonstrates a negative correlation with the source luminosity \citep[see
also][]{cycline} similar that observed in several other bright X-ray
transient pulsars: 4U\,0115+63 \citep{2006ApJ...646.1125N,
2007AstL...33..368T} and V\,0332+53 \citep{2010MNRAS.401.1628T}. The ratio of
fluxes in the 0.5--100 and 0.5--10 keV energy bands depends also on the
luminosity and lies in the range 2.2--2.8, thus
for the following estimations we used the averaged value of 2.5.
All luminosities discussed below were corrected for the
absorption as well.

\begin{table*}
\begin{center}
\caption{{\it Swift}/XRT observations of the source SMC\,X-2} \label{xrt_all}
\tiny{
\begin{tabular}{ccccc}
\hline\hline
Obs Id &  Date     &  Exposure & Flux$^{a}$ & Mode \\
       &  MJD       &  (s)         & ($10^{-10}$ erg s$^{-1}$ cm$^{-2}$) &      \\

\hline
00034073001 &  57289.6471 & 1979 & 4.52$^{+0.22}_{-0.21}$ & WT \\
00034073002 &  57290.9811 & 1656 & 5.08$^{+0.71}_{-0.64}$ & WT \\
00034073003 &  57292.9435 & 1580 & 6.28$^{+0.09}_{-0.09}$ & WT \\
00034073005 &  57294.5720 & 1799 & 6.28$^{+0.12}_{-0.11}$ & WT \\
00034073007 &  57296.3051 & 1947 & 6.31$^{+0.13}_{-0.12}$ & WT \\
00034073008 &  57297.1372 & 1888 & 5.53$^{+0.08}_{-0.08}$ & WT \\
00034073009 &  57298.6319 & 1908 & 5.09$^{+0.07}_{-0.06}$ & WT \\
00034073010 &  57299.7248 & 1760 & 4.91$^{+0.10}_{-0.09}$ & WT \\
00034073011 &  57300.8618 & 491  & 4.27$^{+0.13}_{-0.14}$ & WT \\
00034073012 &  57299.7119 & 405  & 4.76$^{+0.18}_{-0.17}$ & WT \\
00034073013 &  57301.5726 & 224  & 3.15$^{+0.19}_{-0.18}$ & WT \\
00034073014 &  57301.5834 & 1597 & 2.81$^{+0.07}_{-0.08}$ & WT \\
00034073015 &  57302.8351 & 294  & 3.24$^{+0.15}_{-0.15}$ & WT \\
00034073016 &  57302.8475 & 1758 & 3.78$^{+0.10}_{-0.10}$ & WT \\
00034073017 &  57303.3978 & 183  & 2.67$^{+0.22}_{-0.20}$ & WT \\
00034073018 &  57303.4044 & 1965 & 3.70$^{+0.09}_{-0.08}$ & WT \\
00034073019 &  57304.8662 & 301  & 3.17$^{+0.15}_{-0.15}$ & WT \\
00034073020 &  57304.8727 & 1979 & 3.30$^{+0.06}_{-0.05}$ & WT \\
00034073021 &  57305.8641 & 388  & 2.79$^{+0.12}_{-0.11}$ & WT \\
00034073022 &  57305.8710 & 1984 & 3.25$^{+0.05}_{-0.06}$ & WT \\
00034073023 &  57306.7901 & 361  & 2.72$^{+0.13}_{-0.12}$ & WT \\
00034073024 &  57306.7974 & 1991 & 2.94$^{+0.05}_{-0.05}$ & WT \\
00081771001 &  57307.9317 & 471  & 2.28$^{+0.10}_{-0.10}$ & WT \\
00081771002 &  57307.9366 & 1497 & 2.83$^{+0.06}_{-0.06}$ & WT \\
00034073025 &  57308.7849 & 410  & 2.17$^{+0.11}_{-0.11}$ & WT \\
00034073026 &  57308.7924 & 2158 & 2.54$^{+0.05}_{-0.05}$ & WT \\
00034073027 &  57309.3130 & 288  & 2.83$^{+0.15}_{-0.15}$ & WT \\
00034073028 &  57309.3141 & 1005 & 2.59$^{+0.07}_{-0.07}$ & WT \\
00034073029 &  57310.5200 & 348  & 2.17$^{+0.11}_{-0.10}$ & WT \\
00034073030 &  57310.5274 & 2088 & 2.48$^{+0.05}_{-0.05}$ & WT \\
00034073031 &  57311.6110 & 381  & 2.23$^{+0.11}_{-0.11}$ & WT \\
00034073032 &  57311.6184 & 4008 & 2.34$^{+0.04}_{-0.03}$ & WT \\
00034073033 &  57312.4362 & 1358 & 2.07$^{+0.05}_{-0.05}$ & WT \\
00034073034 &  57312.4409 & 4500 & 2.01$^{+0.03}_{-0.03}$ & WT \\
00034073035 &  57313.4667 & 520  & 1.87$^{+0.08}_{-0.08}$ & WT \\
00034073037 &  57314.8068 & 224  & 1.87$^{+0.14}_{-0.13}$ & WT \\
00034073038 &  57314.7121 & 1560 & 1.77$^{+0.05}_{-0.04}$ & WT \\
00034073039 &  57315.2938 & 524  & 1.74$^{+0.09}_{-0.08}$ & WT \\
00034073040 &  57315.3039 & 4355 & 1.55$^{+0.03}_{-0.02}$ & WT \\
00034073041 &  57316.6603 & 521  & 1.68$^{+0.08}_{-0.08}$ & WT \\
00034073042 &  57316.6686 & 3953 & 1.74$^{+0.03}_{-0.03}$ & WT \\
00034073043 &  57317.2878 & 943  & 1.56$^{+0.06}_{-0.06}$ & WT \\
00034073044 &  57317.2969 & 8329 & 1.42$^{+0.02}_{-0.02}$ & WT \\
00034073045 &  57318.3171 & 605  & 1.48$^{+0.09}_{-0.08}$ & WT \\
00034073047 &  57319.4473 & 151  & 1.16$^{+0.21}_{-0.18}$ & WT \\
00034073048 &  57319.4910 & 2228 & 1.25$^{+0.04}_{-0.04}$ & WT \\
00034073049 &  57320.4445 & 391  & 1.21$^{+0.10}_{-0.09}$ & WT \\
00034073051 &  57321.3082 & 496  & 1.35$^{+0.08}_{-0.08}$ & WT \\
00034073052 &  57321.3155 & 3938 & 1.24$^{+0.03}_{-0.03}$ & WT \\
00034073054 &  57322.3206 & 744  & 1.19$^{+0.06}_{-0.06}$ & WT \\
00034073055 &  57323.1040 & 600  & 1.12$^{+0.07}_{-0.06}$ & WT \\
00034073056 &  57323.1123 & 3484 & 1.03$^{+0.03}_{-0.03}$ & WT \\
00034073058 &  57324.6376 & 312  & 1.30$^{+0.16}_{-0.14}$ & PC \\
00034073059 &  57325.1331 & 720  & 1.08$^{+0.07}_{-0.06}$ & WT \\
00034073060 &  57325.1382 & 3678 & 1.15$^{+0.03}_{-0.03}$ & WT \\
00034073061 &  57326.3635 & 778  & 1.01$^{+0.06}_{-0.06}$ & WT \\
00034073062 &  57326.3673 & 3905 & 1.07$^{+0.03}_{-0.03}$ & WT \\

\hline
\end{tabular}
}
\end{center}
\end{table*}

\begin{center}
\tiny{
\begin{tabular}{ccccc}
\hline\hline
Obs Id &  Date     &  Exposure & Flux$^{a}$ & Mode \\
       &  MJD       &  (s)        & ($10^{-10}$ erg s$^{-1}$ cm$^{-2}$) &      \\
\hline
00034073063 &  57327.1952 & 744  & 0.93$^{+0.07}_{-0.04}$ & WT \\
00034073064 &  57327.1999 & 2325 & 1.07$^{+0.04}_{-0.04}$ & WT \\
00034073065 &  57328.1589 & 549  & 0.93$^{+0.07}_{-0.06}$ & WT \\
00034073067 &  57337.9300 & 448  & 0.71$^{+0.07}_{-0.05}$ & WT \\
00034073068 &  57337.9357 & 1389 & 0.91$^{+0.04}_{-0.06}$ & WT \\
00034073070 &  57338.7359 & 1703 & 0.91$^{+0.04}_{-0.02}$ & WT \\
00034073071 &  57339.7532 & 550  & 0.66$^{+0.06}_{-0.04}$ & WT \\
00034073072 &  57339.7619 & 2365 & 0.85$^{+0.04}_{-0.02}$ & WT \\
00034073073 &  57340.8150 & 134  & 0.66$^{+0.13}_{-0.11}$ & WT \\
00034073074 &  57340.7626 & 1750 & 0.76$^{+0.04}_{-0.03}$ & WT \\
00034073075 &  57341.6784 & 360  & 0.60$^{+0.07}_{-0.05}$ & WT \\
00034073076 &  57341.7213 & 2016 & 0.63$^{+0.03}_{-0.03}$ & WT \\
00034073077 &  57342.8066 & 294  & 0.78$^{+0.09}_{-0.07}$ & WT \\
00034073078 &  57342.8219 & 1038 & 0.79$^{+0.04}_{-0.05}$ & WT \\
00034073079 &  57343.6030 & 736  & 0.63$^{+0.06}_{-0.04}$ & WT \\
00034073080 &  57343.6094 & 1777 & 0.74$^{+0.03}_{-0.05}$ & WT \\
00034073081 &  57345.9011 & 178  & 0.41$^{+0.12}_{-0.10}$ & WT \\
00034073082 &  57345.9076 & 1995 & 0.63$^{+0.03}_{-0.03}$ & WT \\
00034073083 &  57346.3602 & 384  & 0.56$^{+0.07}_{-0.05}$ & WT \\
00034073084 &  57346.3676 & 1989 & 0.51$^{+0.04}_{-0.03}$ & WT \\
00034073085 &  57347.6270 & 236  & 0.58$^{+0.10}_{-0.09}$ & WT \\
00034073086 &  57347.6336 & 1922 & 0.52$^{+0.02}_{-0.04}$ & WT \\
00034073087 &  57348.5565 & 170  & 0.49$^{+0.09}_{-0.08}$ & WT \\
00034073088 &  57348.5630 & 1985 & 0.51$^{+0.02}_{-0.02}$ & WT \\
00034073089 &  57349.3533 & 277  & 0.49$^{+0.06}_{-0.06}$ & WT \\
00034073090 &  57349.3554 & 1171 & 0.58$^{+0.04}_{-0.04}$ & WT \\
00034073091 &  57350.4840 & 252  & 0.48$^{+0.07}_{-0.06}$ & WT \\
00034073092 &  57350.4907 & 1703 & 0.52$^{+0.04}_{-0.02}$ & WT \\
00034073093 &  57351.5410 & 240  & 0.50$^{+0.09}_{-0.08}$ & WT \\
00034073094 &  57351.5479 & 2338 & 0.58$^{+0.03}_{-0.04}$ & WT \\
00034073095 &  57352.7369 & 290  & 0.38$^{+0.05}_{-0.05}$ & WT \\
00034073096 &  57352.7434 & 2046 & 0.45$^{+0.03}_{-0.02}$ & WT \\
00034073099 &  57354.5984 & 323  & 0.35$^{+0.05}_{-0.05}$ & WT \\
00034073100 &  57354.6058 & 2232 & 0.42$^{+0.02}_{-0.02}$ & WT \\
00034073102 &  57355.7453 & 2101 & 0.46$^{+0.03}_{-0.03}$ & WT \\
00034073103 &  57356.5266 & 444  & 0.26$^{+0.05}_{-0.04}$ & WT \\
00034073104 &  57356.5342 & 1972 & 0.45$^{+0.03}_{-0.03}$ & WT \\
00034073105 &  57357.1884 & 242  & 0.30$^{+0.04}_{-0.04}$ & WT \\
00034073106 &  57357.1953 & 2165 & 0.41$^{+0.02}_{-0.03}$ & WT \\
00034073108 &  57358.3574 & 977  & 0.35$^{+0.03}_{-0.03}$ & WT \\
00034073109 &  57359.1477 & 273  & 0.22$^{+0.05}_{-0.04}$ & WT \\
00034073110 &  57359.1553 & 1988 & 0.27$^{+0.02}_{-0.02}$ & WT \\
00034073111 &  57360.2485 & 391  & 0.22$^{+0.05}_{-0.04}$ & WT \\
00034073112 &  57360.2559 & 1981 & 0.25$^{+0.02}_{-0.02}$ & WT \\
00034073113 &  57361.3750 & 3266 & 0.23$^{+0.05}_{-0.04}$ & WT \\
00034073114 &  57361.3819 & 1983 & 0.22$^{+0.02}_{-0.02}$ & WT \\
00034073116 &  57362.1417 & 1297 & 0.15$^{+0.02}_{-0.02}$ & WT \\
00034073117 &  57363.1686 & 348  & 0.13$^{+0.05}_{-0.04}$ & WT \\
00034073118 & 57363.1778 & 2104.6 & 0.13$^{+0.02}_{-0.02}$ & WT \\
00034073119 & 57364.3003 & 147.5 & 0.17$^{+0.09}_{-0.06}$ & WT \\
00034073120 & 57364.3073 & 1987.7 & 0.09$^{+0.01}_{-0.01}$ & WT \\
00034073121 & 57365.3633 & 218.3 & 0.08$^{+0.05}_{-0.03}$ & WT \\
00034073122 & 57365.3701 & 1977.6 & 0.11$^{+0.02}_{-0.02}$ & WT \\
00034073123 & 57366.1267 & 175.0 & 0.07$^{+0.08}_{-0.04}$ & WT \\
00034073124 & 57366.0685 & 1111.0 & 0.08$^{+0.02}_{-0.01}$ & WT \\
00034073127 & 57368.4540 & 114.3 & 0.08$^{+0.06}_{-0.04}$ & WT \\
00034073128 & 57368.4602 & 861.8 & 0.07$^{+0.02}_{-0.01}$ & WT \\
00034073129 & 57369.1544 & 368.2 & 0.10$^{+0.04}_{-0.03}$ & WT \\
\hline
\end{tabular}
}
\end{center}

\begin{center}
\tiny{
\begin{tabular}{ccccc}
\hline\hline
Obs Id &  Date     &  Exposure & Flux$^{a}$ & Mode \\
       &  MJD       &  (s)        & ($10^{-10}$ erg s$^{-1}$ cm$^{-2}$) &      \\
\hline

00034073130 & 57369.1618 & 1340.8 & 0.04$^{+0.01}_{-0.01}$ & WT \\
00034073131 & 57370.9811 & 95.7 & 0.07$^{+0.09}_{-0.04}$ & WT \\
00034073132 & 57370.9893 & 1231.4 & 0.03$^{+0.01}_{-0.01}$ & WT \\
00034073133 & 57371.2809 & 221.6 & 0.29$^{+0.13}_{-0.14}$ & WT \\
00034073134 & 57371.2871 & 1782.7 & 0.06$^{+0.02}_{-0.01}$ & WT \\
\hline
\multicolumn{5}{c}{Group 1}\\
00034073135 &  57375.5719 & 1938 &  0.0003$^b$ & PC \\
00034073136 &  57376.6368 & 1583 &         & PC \\
00034073137 &  57378.2005 & 1568 &         & PC \\
\hline
\multicolumn{5}{c}{Group 2}\\
00034073139 &  57392.3539 & 737  &  0.00016$^b$ & PC \\
00034073140 &  57394.9368 & 1653 &  & PC \\
00034073141 &  57396.2917 & 912  &  & PC \\
00034073142 &  57398.4597 & 1586 &  & PC \\
00034073143 &  57400.5460 & 1900 &  & PC \\
00034073144 &  57402.3410 & 1836 &  & PC \\
00034073145 &  57404.6973 & 2379 &  & PC \\
00034073146 &  57406.3691 & 1226 &  & PC \\
00034073147 &  57408.3840 & 1671 &  & PC \\
00034073148 &  57410.7484 & 1848 &  & PC \\
00034073149 &  57412.7004 & 812  &  & PC \\
\hline
\end{tabular}
}
\end{center}

{\tiny
\begin{tabular}{ll}
$^{a}$  & Unabsorbed flux in the 0.5--10 keV energy range\\
$^{b}$  & Upper limit in the 0.5--10 keV energy range \\
\end{tabular}
}
\vspace{5mm}

\begin{figure}
\figurenum{1}
\centerline{
\includegraphics[width=0.6\columnwidth,bb=20 220 570 680,clip]{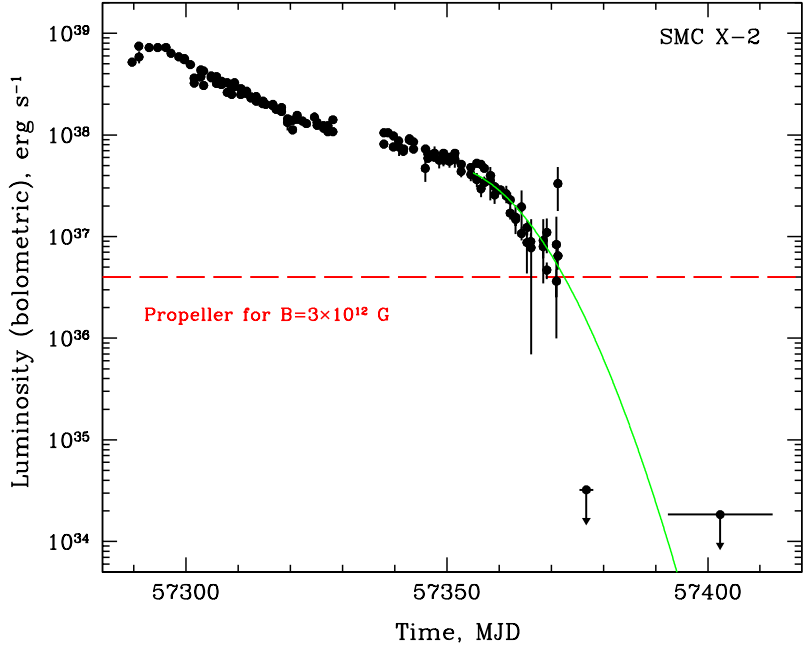}
}
\caption{Evolution of the bolometric luminosity of \smc\ during the
2015--2016 outburst. Luminosity is calculated from the unabsorbed flux
derived from {\it Swift}/XRT data under assumption of the distance $d=62$ kpc
and bolometric correction factor of 2.5 (see Section \ref{sec:res}). Two
upper limits were obtained by averaging 3 and 11 observations, respectively,
with very low count statistic (note that all of them were performed in the PC
mode). Solid green line illustrates the flux decay law right before the
transition to the propeller regime and obtained by fitting of the light curve
with a Gaussian function \citep[see, e.g.,][]{2014MNRAS.441.1984C}.
Horizontal dashed line shows the approximate limiting luminosity when the
propeller regime sets in.}\label{fig:lcxrt}
\end{figure}

The bolometric luminosity of \smc\ corrected for the absorption is presented
in Fig.~\ref{fig:lcxrt}. There are (at least) three interesting features in
this light curve:
\begin{enumerate}
\item The maximum luminosity is about $10^{39}$\,\lum, that exceeds the
standard Eddington limit for the neutron star by a factor of 5. This implies
that at such a high accretion rate, the accretion column should be formed at
the neutron star surface and the magnetic field on the neutron star surface
should be at least $\sim10^{12}$~G \citep{2015MNRAS.454.2539M}.
\item The decay of the light curve is nearly exponential during the first two
months of the outburst (excluding last dozen of days). \item There is a
dramatic drop (by a factor of more than 100) of the source luminosity on MJD
57370 at around $L_{\rm lim}\simeq(4\pm1) \times10^{36}$\,\lum\ (see red
dashed line). Note, that there is a time gap (about 4 days) between the
last observation when the source was significantly detected and the next
one when the source was not detected already (see Table\,\ref{xrt_all}).
Therefore we take the limiting luminosity as the average luminosity between the
last significant measurement and an extrapolation of our fit (see gaussian in
Fig. 1) to the time of the next observation where the source was not
detected. The uncertainty corresponds to the difference between that
luminosity and the measured or extrapolated luminosities. This drastic change
of the source luminosity is most likely related to its transition to the
propeller regime.
\end{enumerate}

Below we briefly discuss this effect and estimate physical parameters of the
system \smc, in particular, the magnetic field strength of the neutron star.

\begin{figure}
\figurenum{2}
\centerline{
\includegraphics[width=0.6\columnwidth,bb=50 270 565 680,clip]{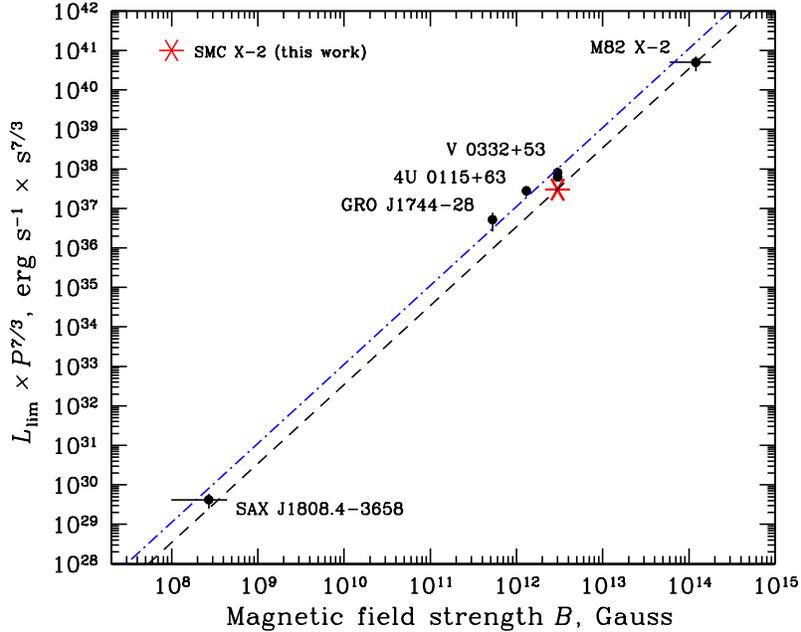}
}
\caption{The observed correlation between a combination of the propeller
limiting luminosity and the pulse period, $L_{\rm lim}P^{7/3}$, and the
magnetic field strength $B$ at the neutron star surface  (independent
measurements) for five sources \citep[for the details
see][]{2016A&A...593A..16T}. Dashed and dot-dashed lines represent the
theoretical dependence from Equation~(\ref{eq1}) assuming standard parameters for
a neutron star ($M=1.4 M_{\odot}$, $R=10$~km) and $k=0.5$ and $k=0.7$,
respectively. The red star indicates the position of \smc\ for the case of
$k=0.5$ and the limiting luminosity $L_{\rm lim}\simeq4\times10^{36}$
\lum.}\label{fig:prop}
\end{figure}

\section{Estimate of the magnetic field}
\label{sec:dis}

The main idea behind the propeller effect is that the accretion of matter
onto a strongly magnetized neutron star is possible only if the velocity of
the magnetic field lines is lower than the local Keplerian velocity at the
magnetospheric radius ($R_{\rm m}$). This condition is fulfilled only in the
case when the magnetospheric radius  is smaller than the co-rotation radius
($R_{\rm c}$). Because of the dependence of the magnetospheric radius on the
mass accretion rate and magnetic field strength, one can link the latter to
the limiting luminosity corresponding to the onset of the propeller regime by
equating the co-rotation and magnetospheric radii \citep[][]{2002ApJ...580..389C}

\be \label{eq1}
L_{\rm lim}(R) \simeq \frac{GM\dot{M}_{\rm lim}}{R} \simeq 4 \times 10^{37} k^{7/2} B_{12}^2
P^{-7/3} M_{1.4}^{-2/3} R_6^5 \,\textrm{erg s$^{-1}$},
\ee
where $R_6$ is neutron star radius in units of $10^6$~cm, $M_{1.4}$ is the
neutron star mass in units of 1.4M$_\odot$, $P$ is the pulsar's rotational period
in seconds, $B_{12}$ is the magnetic field strength in units of $10^{12}$~G
on the neutron star surface under an assumption of the dipole configuration
of the magnetic field. The factor $k$ relates the magnetospheric radius in
the case of disc accretion to the Alfv\'en radius calculated for spherical
accretion ($R_{\rm m}=k \times R_{\rm A}$) and is usually assumed to be
$k=0.5$ \citep{1978ApJ...223L..83G}.

As it follows from Equation~(\ref{eq1}), a detection of the transition of the
pulsar to the propeller regime and the measurement of the corresponding
limiting luminosity $L_{\rm lim}$ can be used to estimate the neutron star
magnetic field. The validity of such approach was recently proven by
\citet{2016A&A...593A..16T} who compared for several X-ray pulsars the
magnetic field strengths obtained with this method with independently
determined values in extremely wide range of magnetic fields, from $10^8$ to
$10^{14}$ G (Fig.\,\ref{fig:prop})

For the standard parameters of a neutron star of $M=1.4 M_{\odot}$, $R=10$~km
and $k=0.5$, the measured limiting luminosity of $L_{\rm lim}\simeq 4 \times
10^{36}$ \lum\ corresponds to the magnetic field strength $B=(3.0\pm0.4)
\times 10^{12}$~G. Based on these estimations we can predict that the
cyclotron line should be observed in \smc\ at around 25~keV. This is in
agreement with the independent measurement by the {\it NuSTAR} observatory
\citep[see above and][]{cycline}.

\section{Conclusion}

In this paper we have reported the discovery of the propeller effect in the
bright transient X-ray pulsar \smc. The dramatic drop of the source
luminosity (by a factor of more than 100) on the time scale of a few days was
revealed thanks to the monitoring campaign with the {\it Swift}/XRT
telescope, organized during the Type II outburst registered from the source
in 2015 September  -- 2016 January. The luminosity drop occurred near the
luminosity of $L_{\rm lim}\simeq4\times10^{36}$~\lum. Based on this
measurement we estimated the magnetic field strength of the neutron star in
the \smc\ binary system as $B\simeq3\times10^{12}$~G, that is typical for
X-ray pulsars \citep[see, e.g., recent review by][]{walter15} and
confirmed independently by results of the spectral analysis. Thus our
discovery makes \smc\ the sixth known pulsating X-ray source where the
propeller effect was observed.

\acknowledgments
This research has made by using {\it Swift}/XRT data provided by the UK Swift
Science Data Centre at the University of Leicester. AAL, SST and SVM
acknowledge support from Russian Science Foundation (grant 14-12-01287). JP
thanks the Foundations' Professor Pool, the Finnish Cultural Foundation and
the Academy of Finland (grant 268740) for financial support. Authors thanks
to the anonymous referee for the useful comments.

\label{lastpage}
\end{document}